# Serverless Computing for Cloud-based Power Grid Emergency Generation Dispatch

Song Zhang, *Senior Member, IEEE*, Xiaochuan Luo, *Senior Member, IEEE*, and Eugene Litvinov, *Fellow, IEEE*

*Abstract*—Operating a modern power grid reliably in case of SCADA/EMS failure or amid difficult times like COVID-19 pandemic is a challenging task for grid operators. In [11], a PMU-based emergency generation dispatch scheme has been proposed to help the system operators with the supply and demand balancing; however, its realization highly relies on the control center infrastructure for computing and communication. This work, rather than using the on-premises server and dispatch communication system, proposes and implements a cloud-centric serverless architecture to ensure the operation continuity regardless of local infrastructure's availability and accessibility. Through its prototype implementation and evaluation at ISO New England, the solution has demonstrated two major advantages. Firstly, the cloud infrastructure is independent and fault-tolerant, providing grid monitoring and control capability even when EMS loses the corresponding functionality or when operators need to work remotely away from the control center. Secondly, the overall design is event-driven using serverless service in response to the SCADA/EMS failure event. Thanks to "serverless", the burden of the server provisioning and maintenance can be avoided from the user side. The cost of using public cloud services for this solution is extremely low since it is architected and implemented based on the event-driven Function-as-a-Service (FaaS) model. This work also develops a comprehensive cyber security mechanism to comply with critical infrastructure requirements for the power grid, which can serve as an exemplary framework for other grid operators to secure their cloud services.

## I. Introduction

POWER grid is a typical cyber-physical system [1], which means that at any time, there is not only the energy flow travelling from one location to another in the network, but also information flow moving across the power grid and control centers. Secure and reliable grid operation requires controlling the energy flow in a manner that supply and demand can be well balanced in real time by employing the information flow. Ensuring the information moves as expected is thus a prerequisite for the sustained operation of power systems under normal conditions. Alternatively speaking, any interruption of information flow would disturb the right behavior of energy flow, and further undermine the system's secure and reliable operation.

To enable real-time monitoring and data acquisition that is essential for the system operation and control, the modern power grid primarily relies on the Supervisory Control and Data Acquisition (SCADA) system, which is tightly coupled with the Energy Management System (EMS) to guarantee the critical information is transferred reliably. After decades of development, the SCADA/EMS has been designed to be fault-tolerant and highly available. The system is usually built with physical redundancy by means of dual servers and software redundancy through in-memory data replication mechanism [2]; however, loss of SCADA/EMS system, either partially or entirely, could still happen from time to time. Statistics show that in the past four years from October 2013 to April 2017, there are 318 such events reported to North America Electric Reliability Corporation (NERC), the regulatory authority in North America who develops and enforces electric power grid reliability standards. In this report, communication interruption was found to be the major factor leading to these system failures [3]. According to a guideline developed by NERC, loss of SCADA or EMS is categorized as an "Emergency", and each Transmission Operator and Balancing Authority (BA) must have operating procedures and plans to mitigate its effects on the grid operation [4].

To balance the power supply and demand for a specific control area in real time, grid operators run a Security Constrained Economic Dispatch (SCED) program at the control center and issue dispatch orders to all dispatchable generators. BAs also rely on Automatic Generation Control (AGC), a critical application in EMS, to send control orders at every four seconds to AGC-participating generators. The inputs to AGC, which come from SCADA system, include measurements of the power flow on the tie-lines that connect two neighboring control areas and frequencies on key buses in the control area. Obviously, loss of SCADA or EMS will paralyze SCED and AGC, further threatening the power balance and frequency regulation. On the other hand, the generator dispatch instructions and AGC control set points are sent through a dedicated communication system that connects the BA and generators, for example, PJM*net* at PJM [5], Automated Dispatch System at California ISO [6], and Electronic Dispatch system at ISO New England [7]. The general data flow of the dispatch instructions from the system operator to the

Song Zhang, Xiaochuan Luo, Eugene Litvinov are with Department of Business Architecture & Technology, ISO New England, Inc., Holyoke, MA 01040, USA (emails: sozhang@iso-ne.com, xluo@iso-ne.com, elitvinov@iso-ne.com)

responsible participant for generator control is illustrated in Fig. 1.

As it can be seen from Fig. 1, a typical dispatch communication system consists of three main parts, the Remote Telemetry Unit (RTU) at each participating unit, a private redundant communication network established by the BA, and the Communication Front End (CFE) Servers at the grid control center. Communication between each RTU and the CFE uses a proprietary protocol over TCP/IP to provide the ability to collect data and distribute supervisory control commands to and from generators in real time. Under normal condition, the generator Desired Dispatch Points (DDPs) are calculated by SCED and AGC at the BA side, passed to the CFE servers and then transmitted to the generator's RTU via the private communication route.

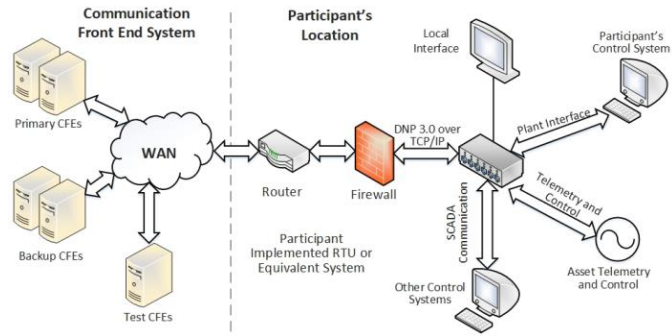

Fig. 1 Data flow of the dispatch instructions from BA to generator

Though the functionality of the automatic generator dispatch could be interrupted, maintaining the real-time supply-demand balance cannot be paused because it is the foundation for the stable operation of entire power systems. Current operating procedures require operators continuing performing "Emergency" dispatch, in the event of loss of SCADA or EMS, by manually sending estimated DDPs to generators via the dispatch communication system [8], which could be inaccurate. Specifically, when the system has large uncertainty and variability with high penetration of renewable energy, manual dispatch using the data from the outdated system snapshot and lagged forecast will become increasingly inaccurate. If the dispatch communication system is unavailable at that moment, operators must communicate with generators via secure phone calls to perform verbal dispatch [7][9]. Such a procedure has significant manual efforts involved, making the system operation performance vulnerable to human errors. Besides, it is also progressively difficult to dispatch multiple generators verbally in a short interval to balance the system.

In addition to critical infrastructure failures, operators may also have to deal with unusual times when the facilities are inaccessible, like COVID-19. The virus outbreak greatly increases the probability of operation disruption because operators lack timely support from IT and operation engineers, who are required to work from home amid the pandemic. An even worse circumstance is operators are infected and home quarantined. Since not all essential applications for generation dispatch allow remote access, how to continue operating the grid during such situations becomes of particular interest to grid operators.

In our previous work [10], a synchrophasor infrastructure based control framework was first proposed to use Phasor Measurement Unit (PMU) data as a backup of SCADA data, when the latter is unavailable, to calculate Area Control Error (ACE) and corresponding generation dispatch amount. Then in the subsequent work [11], an on-premises PMU-based generation control scheme was developed to perform AGC-like dispatch to minimize ACE and economic dispatch to follow the projected load change. The control scheme can maintain system balance in a way that the NERC's balancing control performance requirements can be met even during the "Emergency" period. Although these works have put forward an effective method of grid monitoring and control in the event of SCADA or EMS failure, their realizations still have high dependence on the BA's infrastructure for computing and communication. In other words, once the local infrastructure is unavailable or inaccessible, the control scheme deployed in house will become invalid.

On the other hand, cloud computing has become as vital as power, transport and fresh water supply today. In June 2019, Federal Energy Regulatory Commission (FERC) held a technical conference, which has extensive discussions upon the use of cloud services in the power industry. In the meeting minutes, the utilities and regulators are encouraged to work collaboratively to ensure secure and reliable adoption of cloud technology [12]. Just a few months later, FERC issued a Notice of Inquiry (NOI) to seek comments and suggestions regarding the potential benefits and risks associated with the use of virtualization and cloud computing technologies [13].

Give the background, we present a serverless solution for the realization of the entire emergency generation dispatch based on public cloud services. In this work, we continue to use the PMU data that is independent of SCADA. Different from [11], which developed a PMU-based generation dispatch scheme running in control center to counteract the impact of SCADA/EMS failures, this work presents a brand-new architecture with cloud-hosted data ingestion, processing, storage, visualization and sharing to ensure the operation continuity in case of unavailability and inaccessibility of control center facilities. This cloud-centric architecture has two major advantages over the traditional solution architecture that heavily relies on on-premises infrastructure. Firstly, the cloud infrastructure is independent and fault-tolerant, providing backup grid monitoring and control capability even when EMS loses its corresponding functionality or when operators need to work remotely away from the control center. Secondly, the cloud-centric platform is architected on the event-driven Function-as-a-Service (FaaS) model. This serverless architecture relieves customers from the burden of server management and eliminates the cost of server provisioning. The users only need to pay a very low operating cost for the use of service triggered by an "Emergency" due to its short duration and infrequent occurrence. Another contribution of this work is it proposes and implements an exemplary cyber security framework that comply with the requirements for critical energy infrastructure. This will benefit utilities who are seeking cloud-based solutions for their business use cases.

The remainder of this paper is organized as follows. Section II provides a brief introduction to serverless computing including its concept, features, benefits and use cases. Section III first describes the current operation practice when there is a loss of SCADA or EMS, followed by the revisiting of the PMU based generation dispatch algorithm that is proposed in our prior work to counteract the impact of SCADA/EMS failures, then it points out the challenges to this on-premises control scheme in case of control center infrastructure unavailability or inaccessibility. Section IV presents a cloud-hosted framework for emergency generation dispatch. Next, Section V presents a serverless solution that fits into this cloud framework to securely address the need of data processing, storage, visualization and sharing for emergency dispatch, with a highlight on how to meet the cyber security requirements. A conclusion is given at last in Section VI, respectively.

## II. UNDERSTANDING SERVERLESS COMPUTING

Serverless computing is a cloud computing model aiming to abstract server management and low-level infrastructure decisions away from developers [14]. It provides a real pay-as-you-go service with no waste of resources and lowers the bar for developers by entrusting cloud provider to tackle all their operational complexities [15]. Compared to other cloud computing models, serverless computing is closer to original expectations for cloud computing to be treated like a utility service [16]. It is emerging as a new and compelling paradigm for the deployment of cloud applications, mostly due to the recent shift of enterprise application architectures to containers and microservices [17].

The conceptual hierarchy of the serverless architecture is shown in Fig. 2 with a comparison to the traditional Infrastructure-as-a-Service (IaaS) model [18].

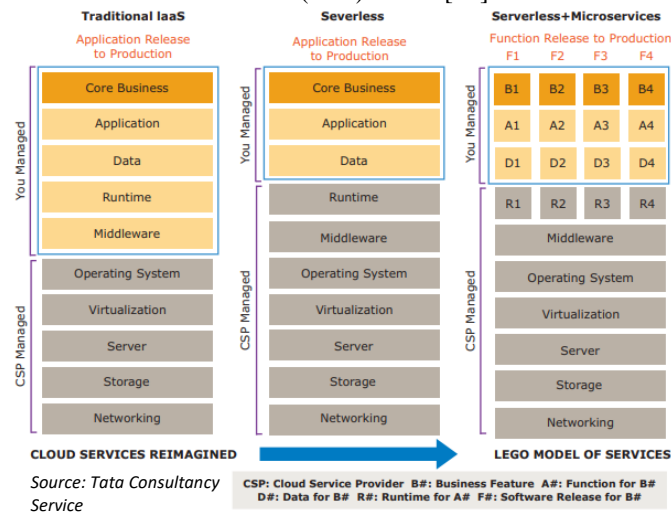

Fig. 2 Traditional IaaS infrastructure and serverless architecture

In this figure, the business features, which are represented by block $Bk$ ($k = 1 \sim 4$), are built on a group of functions. Each function is single-purposed and only does one task. It is comprised of a module from the data layer, which is illustrated by block $Dk$ ($k = 1 \sim 4$), to ingest data, and a corresponding module from the application layer, indicated by block $Ak$ ($k = 1 \sim 4$), to carry out necessary calculation as "function". These functions have their independent runtimes, i.e., block $Rk$ ($k = 1 \sim 4$). Each combination of blocks $Ak$, $Dk$ and $Rk$ respectively constitutes a software release $Fk$ for business feature $Bk$. In the serverless model, users will no longer have to invest in creating, maintaining, or managing the underlying infrastructure. They only need to focus on the development of applications with their business logics embedded. Establishing the business service on the serverless architecture is just like erecting a building with Lego bricks.

### A. Function-as-a-Service

Function-as-a-Service (FaaS) is an implementation of "serverless computing" via serverless architecture. It has become dominating in serverless realization [19] and therefore widely regarded as a synonym of "serverless computing" [20][21]. Because of this, we are not going to differentiate them from each other and will use these two terms interchangeably.

FaaS frees developers from the heavy lifting of building out or maintaining a complex infrastructure by executing code in response to events, which means you can simply upload modular chunks of functionality to the cloud that are executed independently to deploy your service. It was first made available on large commercial cloud platforms by Amazon, which is called AWS Lambda [22]. Following the lead of AWS Lambda, corresponding services such as Azure Functions, Google Cloud Functions, OpenWhisk are introduced by Microsoft, Google, IBM respectively to enrich the options for FaaS implementation [17].

### B. Features and Benefits

Serverless computing, or FaaS, is usually characterized by a number of attributes listed below [17][18].

*1) Event-driven*

FaaS executes the business logic in response to user requests. Anything that triggers the execution of the function is regarded as an event, e.g., message publishing, file upload.

*2) Low cost*

Serverless is a new way of offloading IT overhead. A serverless architecture eliminates the responsibility of managing servers, databases, and even application logic, lowering the setup and maintenance costs. The usage is metered and users pay only for the time and resources used when serverless functions are running. They never need to pay for the idle resources.

*3) Inherently scalable*

Like other cloud computing models, serverless offers inherent scalability. It can not only respond quickly to scale up when needed, but also scale all the way down to zero resources, depending on the user demand. The ability to scale down to zero instance is one of the key differentiators of a serverless platform.

*4) Business focused and productivity centric*

Serverless allows developers to focus on business logic instead of infrastructure so they can improve their productivity of turning ideas to real products or services.

*5) Built-in availability and fault tolerance*

Disaster recovery is integrated into Cloud Service Provider (CSP) offerings. Your applications are highly available on serverless architecture because they are built on fault-tolerant functions.

*6) Stateless*

Functions are stateless because they do not rely on internal memory, state-machines, stored files or mounted volumes. Subject to external services each call to a function should result in the same end-result.

*C. Use Cases*

Serverless computing has been utilized to support a wide range of applications including event processing, stream processing, Extract-Transform-Load (ETL), advanced analytics, Internet of Things (IoT) solutions, artificial intelligence solutions, web and mobile applications, etc.

## III. EMERGENCY GENERATION DISPATCH IN CASE OF LOSS OF EMS OR SCADA

As mentioned in the introduction part as well as in the preceding work [11], system operators are required to perform dispatch using manually collected data when there is an EMS or SCADA failure as pursuant to the emergency operating guidelines by NERC and a variety of BAs. Although these guidelines vary from each other in specific procedures, all of them require significant human intervention during the dispatch process.

*A. Current Operating Practice*

The manual dispatch guidelines by different BAs share some common points. For a grid operator like ISO New England, a typical operating procedure in the event of loss of SCADA/EMS includes the following action steps.

*Action Step 1*: have Local Control Centers (LCCs) deploy personnel staffing key substations; asking for a periodic update of meter reading and topology changes via verbal conversation.

*Action Step 2*: speak to the generator operators to find out the Unit Control Mode (UCM) of the candidate units that may participate in emergency dispatch. A UCM describes the current operational state of a generator, for example, if it is available for dispatch; if it is being postured or regulating. Select generators which are online and available for the dispatch as the participating units for the emergency dispatch.

*Action Step 3*: determine the economic dispatch of the unit and estimate the amount of generation output changes to minimize current ACE and balance projected load change.

*Action Step 4*: send the estimated DDPs to participating units either verbally via secure phone calls, or electronically through the dispatch communication system, if it is still available.

*Action Step 5*: verify that the responsible units have acknowledged the DDPs that are sent to them.

*Action Step 6*: record and log all verbal communications with LCCs, neighboring BAs, and generators.

*B. Economic Generation Dispatch using PMU Measurements*

To reduce the manual effort involved in the current operating procedure during the emergency of SCADA or EMS failure, the PMU measurements can be used as a backup of SCADA to re-acquire the ability of generator dispatch and monitoring. Because the synchrophasor infrastructure is independent of the traditional SCADA system, a failure of SCADA does not necessarily indicate PMU unavailability. Furthermore, unlike the SCADA data, PMU measurements are time-aligned and can be streamed to the cloud either from the substations directly or from the ISO irrespective of network latency. The PMU-based control scheme that proposed in [11] was designed specifically to meet the goal of emergency generation dispatch. With the primary objective of balancing the supply and demand using a cost optimization strategy, the PMU based emergency generation control scheme dispatches generation in response to variation in ACE and projected load change in a short-term look-ahead horizon. The operating limits and ramp rates of dispatchable PMU-monitored generators are respected. Besides, a feasibility check is also enforced to guarantee the LP problem does not end up with an infeasible solution. The details of algorithm including mathematical equations can be found in [11]. This dispatch scheme, which was validated on ISO New England system through near real-time closed loop simulations, demonstrates good balancing control performance that complies with NERC's reliability standard under the emergency when there is a complete loss of SCADA or EMS.

*C. Challenges due to Unavailability and Inaccessibility of the Control Center Infrastructure*

Dispatching PMU-monitored units to maintain system balancing during loss of SCADA or EMS emergency can alleviate the manual workload and reduce operators' mistakes; however, it only provides help to eliminate the human error in action steps 1 and 3 mentioned above. The other action steps, i.e., 2, 4, 5 and 6, are still inevitable, especially when some parts of the control center infrastructure become unavailable, e.g., the dispatch communication system fails. As a matter of fact, the manual dispatch procedure will be triggered and step 2, 4, 5, 6 will be needed whenever there is a failure of the dispatch communication network regardless of the availability of SCADA and EMS. This makes whatever generation dispatch solution ineffective due to the lack of alternative automated communication between grid operators and generators, while depending on phone conversations to issue dispatch instructions is subject to verbal mistakes.

Even though we assume all dispatch instructions over the phone are clearly received and correctly followed by the responsible units, grid operators are unable to dispatch all participating generators simultaneously by verbal conversations. They have to call each unit one by one to communicate the dispatch orders. As a result, the control may be stochastically lagged, uncorrelated and unfavorably offset due to delay in the human reaction. Moreover, solely relying on verbal conversations to acknowledge DDPs is unreliable and slow, which could have an adverse impact on the decision-making for the subsequent dispatch intervals.

The current operation practice using manual verbal dispatch under emergency conditions cannot guarantee the control performance and could put system reliability in jeopardy. It is also highly dependent on the control center infrastructure. Besides, there are situations when grid operators must work remotely away from the control room due to fire, gas leak and mandatory quarantine amid COVID-19, making on-premises facilities inaccessible. It is, therefore, worthwhile to develop an alternative emergency backup solution, which is independent of today's control center infrastructure for computing and communication, to assist operators with system balancing.

## IV. CLOUD-HOSTED FRAMEWORK FOR EMERGENCY GENERATION DISPATCH

The synchrophasor infrastructure can be used in lieu of SCADA system to reinstate the capability of data acquisition and system monitoring for utilities under "Emergency" situations [10]. The generation control algorithm based on the synchrophasor infrastructure can produce dispatch orders for system balancing while minimizing cost [11]. In simple words, the prior works have answered the questions on "how to take in data automatically and fast" and "how to leverage these data for emergency dispatch"; however, the additional question "how to transfer data to other parties quickly and accurately", especially when the control center infrastructure is unavailable, still remains unsolved.

To address this concern, we present a cloud-centric framework as shown in Fig. 3.

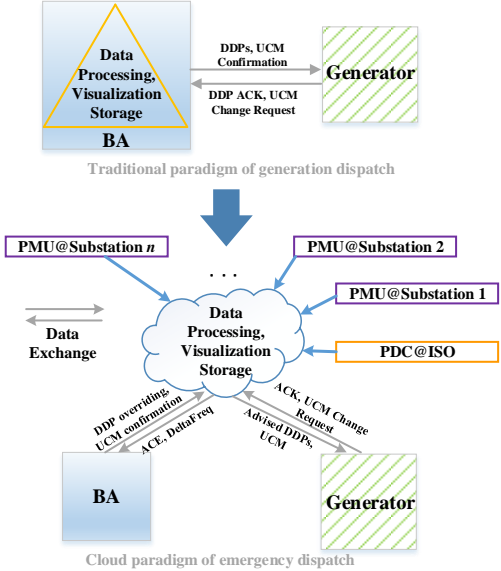

Fig. 3 Transformation from on-premises to the cloud for emergency generation dispatch

The framework transforms the center of information processing, visualization and storage from the on-premises infrastructure to the cloud. The PMU data can either be streamed directly from the substations, or from the Phasor Data Concentrator (PDC) at the BA, to the cloud. The PMU-based control scheme is implemented in the cloud to perform emergency generation dispatch. This cloud framework also enables bi-directional information flow between the BA and generators via secured Internet connections. One of the major benefits is robustness of the data delivery and flexibility of accessing dispatch information from anywhere as long as there is an Internet connection. In a certain sense, this is a sort of Digital Twin [23] in the cloud.

## V. SERVERLESS-BASED EMERGENCY DISPATCH SOLUTION

To realize the cloud-hosted emergency generation dispatch framework, we present a serverless solution with its implementation on Amazon Web Services (AWS) described in detail.

### A. Why Serverless

Before FaaS emerges, a number of other cloud computing models have existed, e.g., Infrastructure-as-a-Service (IaaS), Platform-as-a-Service (PaaS) and Software-as-a-Service (SaaS). Architecting a solution for business on the cloud does not necessarily mean it must be built serverlessly; however, there are some important factors driving us to choose "serverless" or FaaS model in this case.

First of all, an "Emergency" such as loss of SCADA/EMS, failure of dispatch communication network and working remotely away from the control room is an event that occurs infrequently. Generation dispatch for system balancing under the "Emergency" scenario is therefore an event-driven activity. Unlike normal system operation, emergency dispatch is not needed on a 24 x 7 basis. There is no need to provision resources such as servers, databases and I/O capacity for it during normal conditions. Using serverless will help us easily scale down to zero resource consumption when no "Emergency" happens and scale up when emergency dispatch is triggered, at a very low cost. Secondly, the solution for the generation dispatch in the event of an "Emergency" should be made as a group of microservices in order to minimize the impact of any component (subtask) failure in the process. For example, failing to calculate the newest DDPs due to lack of refreshed data will not influence the determination of units' UCM or acknowledgement of previously given dispatch instructions. The FaaS model, which provides the fine-grained services, is the right choice to utilize in this case. Last but not least, we want to avoid the burden of the infrastructure management and maintenance because increasing the IT workload for occasional use case is really undesirable.

### B. Implementation of Serverless Emergency Dispatch

The proposed serverless solution for emergency generation dispatch can be built using any large public cloud service provider like Amazon, Microsoft and Google, because they all have corresponding services for serverless computing that is seamlessly integrated with other services. Having years of experience on running a production-scale cloud platform for power system planning studies on AWS [24], we continued to implement the proposed serverless solution on AWS. As shown in Fig. 4, the entire platform is built upon AWS with its serverless service – Amazon Lambda as the cornerstone. Surrounding Amazon Lambda, the implemented platform also integrates other Amazon services for data ingestion, storage and visualization.

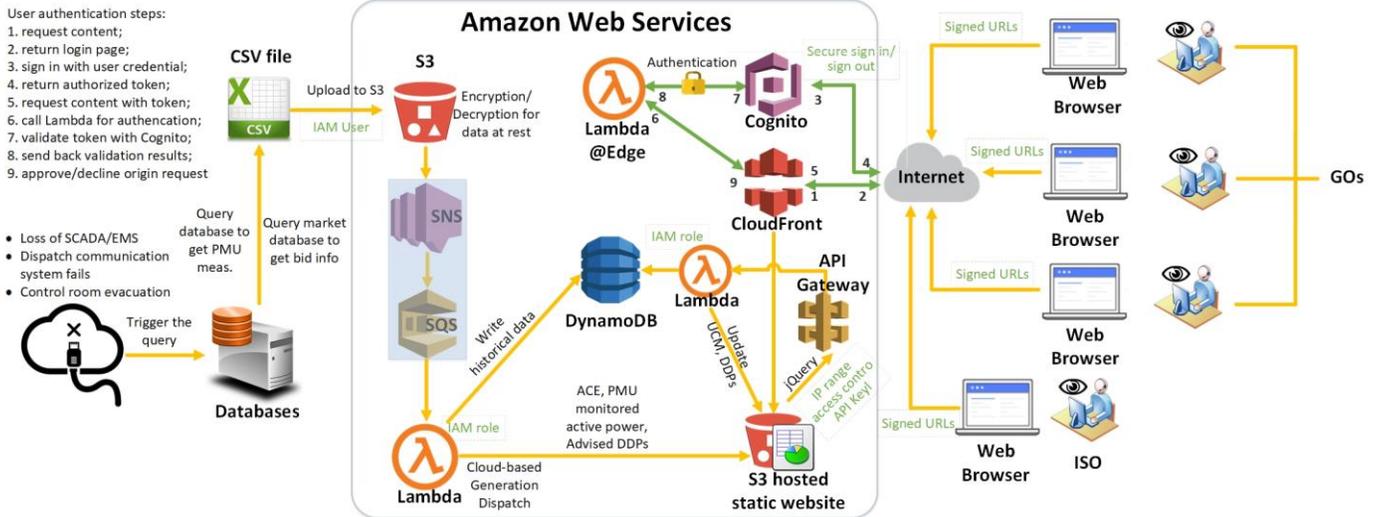

Fig. 4 Serverless implementation of emergency generation dispatch

The implementation of the serverless emergency generation dispatch is described as follows. Whenever there is a loss of SCADA/EMS, dispatch communication system failure or control room evacuation, operators are required to send a notification according to the current operating procedure [7]. This notification is used to trigger a periodic query at every 1 minute to pull the latest (within 1 minute) synchrophasor data, e.g., tie-line flows, frequencies on key buses, and active power outputs of the PMU-monitored generators, from the PMU database. The query also pulls generator offer parameters such as unit ramp rates, incremental energy offers, operating limits and regulation limits from the market database. The data are then wrapped up in a flat file (e.g., .csv) and uploaded to an S3 bucket for encrypted storage (S3 is a scalable, high-speed, low cost, web-based object storage service by Amazon [25]). The activity of putting a data object in the S3 bucket then triggers a Lambda function to retrieve the encrypted data file from the S3 bucket. Once data is parsed from the uploaded file, the Lambda function start calculating ACE and DDPs using the dispatch algorithm described in Section III.B. The calculation results, including ACE, frequency deviation, advised DDPs, are written to a flat file and saved in another S3 bucket where a static website is hosted using Data-Driven Document display technology (d3.js) [26]. In the meanwhile, the historical results are written to a NoSQL database service on AWS, DynamoDB, for archiving purposes.

Alternatively, if the PMU data are directly streamed from the substations to the cloud from substations or from Phasor Data Concentrator (PDC) servers hosted at the BA through Amazon Direct Connect (a secure, dedicated connection from on-premises infrastructure into AWS) as shown in Fig. 3, Amazon Kinesis [27] can be used to ingest the PMU data and deliver them to S3, which will trigger the Lambda function to perform emergency generation dispatch. Currently, we are still working with the cloud provider to establish the dedicated connection to Amazon data centers.

The operators at the BA and the generator side will open their respective webpages to visualize the dispatch information through secure HTTP links in terms of signed Uniform Resource Locators (URLs) after they authenticate themselves with a pre-defined user pool in Amazon Cognito [28]. The operators at the BA side are allowed to manually override the advised DDPs calculated by the Lambda function if they think these values are unreasonable, whereas the generator operators are able to acknowledge the advised DDPs or give a reason to decline it. Similarly, generator operators can initiate a request to change the UCM of a unit and BA can confirm this change. These user requests are completed by API calls through Amazon API Gateway [29]. All the operation activities are logged in DynamoDB database for auditing and responsibility-tracking purpose as a replacement of phone conversation recording.

Amazon S3 service is used together with Amazon SNS, the notification service, and Amazon SQS, the message queuing service, as shown in the shaded part of Fig. 4. Instead of directly using S3 bucket file upload (object putting) events to trigger the Lambda function, the events are sent to a message queue. The Lambda function asynchronously processes the timestamped files in the S3 bucket by looking up the events in the queue. Such a design is helpful if the data files are uploaded at a much smaller interval than the time that Lambda function could process.

*C. Data Visualization*

One benefit of this cloud-centric platform is that it provides a shared platform for consistent data presentation, despite that grid operators and generators need to observe different portion of data due to their different roles in the emergency dispatch process. For BA operators, they are mainly monitoring information related to system balancing performance e.g., ACE, frequency deviation, and Balancing Authority ACE Limit (BAAL) [30], as well as dispatch-related information such as UCM, DDPs and actual outputs of all participating generators. For generators, they only care about the dispatch information for their responsible units, so only that portion of data is displayed to them.

The web-based displays are completely driven by the portion of data that each role desires to view using d3.js libraries. These displays are embedded in serverless interactive webpages created using Hypertext Markup Language (HTML) and JavaScript so that S3 can host them. To enable user requested information sharing, jQuery ajax method is used to send data generated by user actions to web APIs through Amazon API Gateway. The API calls trigger another Lambda which updates the data files in the website hosting S3 bucket and also writes these changes to DynamoDB. Amazon CloudFront service [31] is deployed in the front of S3 hosted website to improve content delivery performance.

*D. Comprehensive Cyber Security Solution*

Cyber security on the cloud is always a top concern. Although cloud computing has been widely used in many areas such as finance, healthcare, insurance, business analytics and even government agencies, it is still an emerging technology to the electric energy industry. Power system users do not typically understand their responsibility on the cloud. They may misplace their trust in the security offerings of the cloud service provider and fail to recognize the additional processes and configurations that need to be in place within their organization to fully secure the solution [32]. In short, lack of understanding in the cloud technology and lack of confidence in cloud security are the major challenges of applying cloud solutions, including the serverless architecture proposed in this work, for power grid operation. To address such concerns, we developed a comprehensive cyber security mechanism, which includes data protection, identity management, key rotation and access authorization as per the instructions given in [33], to secure our cloud solution. It can serve as a good example of security scheme for power system users.

*1) Data Protection*

To protect the Critical Energy Infrastructure Information (CEII), it is mandatory to ensure the data privacy through protecting the sensitive data from being accessed by the malicious or negligent parties. In the cloud context, data protection refers to protecting data both in transit (as it travels to and from Amazon S3), and at rest (while it is stored in Amazon data centers).

To protect data in transit, we enforce the standard Secure Socket Layer (SSL) encryption by using HTTPS protocol. As for protection of data at rest, we opt for the server-side other than client-side encryption because it saves us from the heavy lifting in the encryption/decryption process. There are three options for server-side encryption approach: 1, Use server-side encryption with Amazon S3 managed keys (SSE-S3); 2, use server-side encryption with AWS Key Management Service (KMS) managed keys (SSE-KMS); and 3, use server-side encryption with customer-provided keys (SSE-C). Option 1 requires the minimum effort on key management and encryption process. The key used to encrypt the data is encrypted with a periodically-rotated master key provided by Amazon S3. However, according to [34], the electric energy regulatory bodies tend to use their own encryption key rather than the cloud provider's. Option 3 allows the users to encrypt/decrypt the data by using their own keys, but it will make the data retrieval inefficient because the same key needs to be uploaded each time for data decryption. Based on all considerations above as a whole, we decided to use a mixed mode of option 2 and option 3 in this design with the encryption key generated in-house and managed by the BA while imported to Amazon KMS so that it can be used by other AWS services with ease. For the security concerns, the imported key will be rotated periodically by reimporting the key material. The proposed encryption/decryption process is shown in Fig. 5.

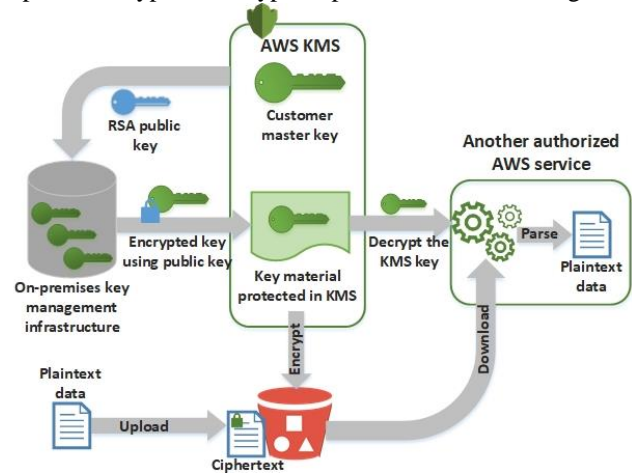

Fig. 5 Workflow of data encryption/decryption

*2) Access Security Check*

In addition to data protection, the access security is also of great importance. All the unauthorized access, no matter visits from outside of AWS infrastructure or requests from one AWS service to another, are prohibited. The locations of green text in Fig. 4 have pointed out the places of access security checkpoints in the implementation. The security check mechanisms include Identity Access Management (IAM) user (a permanent long-term credential for authorized user) for uploading the pulled data to S3, IAM role (an entity that defines a set of permissions for making AWS service requests) for Lambda functions interact with other AWS services, IP range control for limiting the cyber locations of workstations to access the generation dispatch results in S3, API key for sending user requests to API Gateway to acknowledge DDPs or confirm UCM change, signed URLs for serving the private content with sensitive data, and authentication for the viewer's identity through the Lambda@Edge function triggered by CloudFront. The authentication process, which is marked by green double arrows and attached numbers in Fig. 4, include 9 steps.

1). User makes a request for private content
2). CloudFront returns the login page
3). User signs in with his/her credentials
4). Cognito returns authorized token
5). User makes a second request for content with the token
6). CloudFront triggers Lambda for identity authentication
7). Lambda validates the token against Cognito
8). Cognito sends back the validation result, i.e., pass or fail
9). Lambda approves/declines the request

Moreover, using CloudFront to serve HTTPS requests for the S3 hosted website can also absorb cyber-attacks like

Distributed Denial-of-Service (DDoS) attacks.

*E. Cost*

As emergency dispatch is event-driven, using serverless will just incur a very low cost. We ran the emergency dispatch for an hour on AWS. The incurred cost surrounding serverless functions on this platform is given in Table I.

TABLE I
TAXONOMY OF ESTIMATED COST FOR THE USE OF AWS SERVICES

| Service Name | Pricing | Cost ($) |
|---|---|---|
| Lambda | $0.0000002 per request, $0.0000166667 for every GB-SECOND used | ≈ 0.001 |
| S3 | $0.000005 per POST request, $0.0000004 per GET request, $0.023 per GB/month for storage | ≈ 0.00095 |
| DynamoDB | $0.00000125 per write request, $0.00000025 per read request | ≈ 0.000325 |
| CloudFront | $0.0075 per 10,000 HTTP requests | ≈ 0.00018 |
| Others services, e.g., API Gateway, KMS, Cognito | Free tier or negligibly low due to infrequent use | |
| Total | | ≈ 0.0025 |

As analyzed in Table I, the cost of using AWS services for serverless emergency dispatch is literally negligible. By contrast, provisioning local servers in on-premises infrastructure or spinning up virtual machines through traditional IaaS cloud model will lead to inevitable upfront and/or operating costs.

*F. Resource Configuration for Serverless Function Cost-Performance Balance*

Changing the resource settings by selecting memory or changing the timeout may impact the Lambda function performance and cost. Generally speaking, configuration with larger memory or smaller timeout is likely to result in a better performance but higher cost, while choosing smaller memory or larger timeout may yield the opposite combination of metrics. In this use case, the emergency generation control is a lightweight algorithm that requires no significant amount of memory or CPU time. Besides, determined by the characteristics of emergency dispatch, the data do not have to be pulled and uploaded very often. A reasonable upload interval is between 1 minute and 5 minutes. For example, the time used for the Lambda function execution typically ranges between 400 ms and 800 ms, which is much smaller than the function invocation interval, when it is configured with an economy size of memory at 512 MB. The corresponding cost for each request is only 0.5 * $0.0000166667 * mean (0.4, 0.8) + $0.0000002 = $0.000005

*G. Creation and Management of Signed URLs*

Signed URLs are used to restrict access to documents, business data or content that is intended for selected users. They are used in this implementation to block unauthorized access together with user authentication because the data generated by Lambda and stored in S3 are CEII data. The signed URLs are created using a custom CloudFront policy with the expiration date and time specified. The signature can be created in house by the IT department of BA using Base64-encoding tool like OpenSSL [35]. Once created, the signed URLs can be sent to each power plant's point of contact via encrypted email periodically, e.g., once a week or once every other week, to rotate the expired hyperlink.

*H. Deployment for Evaluation at ISO New England*

This serverless-based emergency generation dispatch solution was deployed on AWS and evaluated by ISO New England amid COVID-19 pandemic. It ran in parallel to the regular system balancing process and compared to the latter on several key metrics including ACE, CPS1 and BAAL [28]. Since these data are CEII, we do not disclose them in this work. However, a simplified version of the serverless emergency dispatch with the web-based user interface is available for viewing and playing at http://d3s0lgjz0l0m1d.cloudfront.net/. For simplicity, the user authentication and URL signing process are removed on purpose.

VI. CONCLUSION

The event of loss of SCADA or EMS is usually regarded as an "emergency" by grid operators. To maintain supply-demand balance in case of an emergency, system operators are required to perform manual dispatch by the current operating procedures.

This work presents cloud-hosted serverless solution for emergency generation dispatch using PMU measurements. The solution can not only run economic dispatch to maintain system balance, but also enable rapid, accurate and secure sharing of emergency dispatch information such as control signals, system balancing performance metrics and operator acknowledgements. With an aid of this cloud-based solution, the potential human errors due to the use of phone conversations for manual verbal dispatch can be avoided. Each operation activity done through the platform is written to a database in a clearly defined data format which makes auditing and responsibility tracking much easier. More importantly, the solution also enables the operators to monitor and control the grid even when they need to work remotely away from the control room as long as they have access to the Internet connection.

A comprehensive cyber security mechanism, including data protection, identity management, key rotation and access authorization, was also developed in this work to comply with critical infrastructure requirements for the power grid. This security framework was created in the light of the vendor's official guide to support compliance with NERC CIP standards, which provides a reference for other grid operators to secure their cloud services.

BIOGRAPHIES

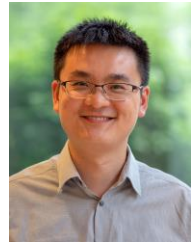
**Song Zhang** (S'11, M'14, SM'20) is a Senior Analyst at ISO New England and a senior member of the IEEE. He is the lead of multiple cloud projects at ISO New England. He is also the Chair of IEEE Power & Energy Society Springfield Chapter. He was a Power System Engineer at GE Grid Solutions from 2014 to 2017. Dr. Zhang received his Ph.D. degree in Electrical Engineering from Arizona State University. His research interest includes power system operation, power system stability and control, cloud computing, big data and synchrophasor technology.

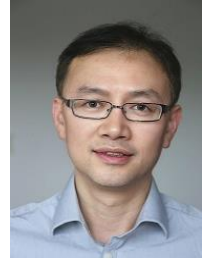
**Xiaochuan Luo** (M'00, SM'15) is a Technical Manager at ISO New England and a senior member of the IEEE. He is responsible for the technology strategy, research and development in power system planning and operations at the ISO. He is the Vice Chair of the IEEE PES Technologies and Innovation Subcommittee. Dr. Luo received his Ph.D. degree in Electrical Engineering from Texas A&M University.

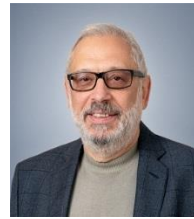
**Eugene Litvinov** (F'13) is the Chief Technologist at ISO New England. He is an IEEE Fellow and a member of National Academy of Engineering. Dr. Litvinov holds BS, MS and Ph.D. in Electrical Engineering. He has more than 40 years of professional experience in the area of power system modeling, analysis and operation; electricity markets design, implementation and operation; information technology.